\documentclass[twocolumn,prl]{revtex4}  % uncomment, if double spacing is 
\usepackage{graphicx}% Include figure files
\usepackage{dcolumn}% Align table columns on decimal point
\usepackage{bm}% bold maths

\begin{document}

\bibliographystyle{apsrev}

\preprint{FIR absorption in disordered systems... }  
  
\title{Universal Features of Terahertz Absorption in Disordered Materials}

\author{S.N. Taraskin}  
 \email{snt1000@cam.ac.uk}  
\affiliation{St. Catharine's College and Dept. of Chemistry, University of Cambridge,  Cambridge, UK}  

\author{S.I. Simdyankin}  
%\email{sis24@cam.ac.uk}  
\affiliation{Dept. of Chemistry, University of Cambridge, Cambridge, UK}  

\author{S.R. Elliott}  
%\email{sre1@cam.ac.uk}  
\affiliation{Dept. of Chemistry, University of Cambridge, Cambridge, UK}  

\author{J.R. Neilson}  
%\email{???}  
\affiliation{Dept. of Chemistry, University of Cambridge,  
             Cambridge, UK and Dept. of Materials Science and 
Engineering, Lehigh University, Bethlehem, P.A. 18015, U.S.A.}  

\author{T. Lo}  
%\email{???}  
\affiliation{Teraview Ltd., Platinum Building, St. John's Innovation Park, 
Cambridge CB4 OW5, UK}

\date{\today}% It is always \today, today,  
  
\begin{abstract} 
Using an analytical theory, experimental terahertz time-domain spectroscopy 
data and numerical evidence, we
demonstrate that the frequency dependence of the absorption coupling
coefficient between far-infrared photons and atomic vibrations in disordered
materials has the universal functional form, 
$C(\omega)=A+B\omega^2$, where the
material-specific constants A and B are related to the distributions 
of fluctuating charges obeying global and local charge
neutrality, respectively. 
\end{abstract}  
  
%  63.50.+x     Vibrational states in disordered systems
%  78.30.Ly     Disordered solids
%  61.43.Fs     Glasses

  
\maketitle  
  
%-------------------------------------------------------------------------  
%                       Main text  
%-------------------------------------------------------------------------  

Since their name indicates a lack of order, it is perhaps unexpected that
all disordered materials, regardless of their chemical composition, often
behave very similarly in response to external probes.
Low-temperature anomalies in thermodynamical properties
of amorphous solids is one example \cite{Phillips_81:book}.
Related to these
is the pile-up of vibrational modes in excess of the Debye vibrational
density of states (VDOS), $g(\omega)\propto \omega^2$, at typical
frequencies $\sim 30~\text{cm}^{-1}$, known as the boson peak (BP)
(see e.g. \cite{Chumakov_04} and Refs. therein).
Interaction of photons with
atomic vibrations in disordered materials is also expected to manifest
universal behaviour, especially in the far-infrared (FIR) frequency domain
\cite{Strom_77}, $1 - 100~\text{cm}^{-1}$, where the atomic vibrations are
material-independent and resemble distorted sound waves. 

The simplest first-order perturbation process of photon interaction with
atomic vibrations is FIR absorption characterized by a linear absorption
coefficient, $\alpha(\omega) = C(\omega)g(\omega)$, which is measured by IR
spectroscopy~--- one of the most valuable experimental techniques to assess
the vibrational properties of amorphous materials
\cite{Galeener_83,Pasquarello_97}~--- and is found generally be
$\alpha(\omega) \propto \omega^2$  in the FIR domain.
The coefficient
$C(\omega)$ quantifies the degree of coupling between IR photons and
atomic vibrations.
It depends on the vibrational eigenmodes 
and the distribution of atomic charges. 
The answer to the question about a possible universal frequency dependence 
of $C(\omega)$ for all or most disordered materials  in the FIR range 
has so far been uncertain, because: 
(i) $C(\omega)$ cannot be measured directly, and two independent
experiments (for measuring $\alpha(\omega)$ and $g(\omega)$) are necessary;
(ii) there is no rigorous analytical theory for $\alpha(\omega)$ in
disordered systems~--- previous theoretical models concentrated mainly on
the
origin of the BP modes \cite{Deich_94} within a particular soft-potential
model or on the role of charge disorder in simple lattice models without
positional disorder  \cite{Schlomann_64}.

In this Letter, we demonstrate that,
for disordered materials in the FIR regime, $C(\omega)$ indeed  exhibits a 
universal frequency dependence of the following form,
\begin{equation}
C(\omega) \simeq A +B \omega^2
~,
\label{e1}
\end{equation}
where $A$ and $B$ are material-dependent constants.
This law is valid in the low-frequency part of the FIR domain, namely
for frequencies below the Ioffe-Regel crossover,
$\omega \alt \omega_{\text{IR}}$,  separating well propagating
plane waves from strongly damped ones due to disorder-induced scattering
\cite{Taraskin_00:PRB_IR1}.
In many materials, the Ioffe-Regel frequency is close to the BP
\cite{Ruffle_06}.
We provide simple
analytical arguments, confirmed by numerical analyses of molecular-dynamics
(MD) models, and terahertz (THz) time-domain spectroscopy (TDS) measurements
of $\alpha(\omega)$ for two glassy systems, SiO$_2$ and As$_2$S$_3$, and use
independently available experimental VDOS data for g-SiO$_2$
\cite{Fabiani_05} and g-As$_2$S$_3$ \cite{Isakov_93,Sokolov_93} to extract
$C(\omega)$.

Our analytical arguments are based on an  analysis of 
the following general expression for the coefficient of absorption  of FIR 
photons by harmonic atomic vibrations obtained within the 
rigid-ion model \cite{Maradudin_61}, 
\begin{equation}
C(\omega) \simeq 
C_0 
\left| 
\sum_i \frac{q_i}{\sqrt{m_i}}{\bf e}_{i}(\omega)
\right|^2 
~,    
\label{e2} 
\end{equation}
where $C_0=2\pi^2n/c\varepsilon^{1/2}_{\infty}$, with  
$m_i$ and $q_i$ being the atomic masses and fixed atomic charges, 
${\bf e}_{i}(\omega)$ the component of the 
eigenvector of frequency $\omega$ corresponding to atom $i$, 
$\varepsilon_{\infty}$ stands for the high-frequency dielectric constant 
and $n$ is the atomic concentration. 
A more general model with the fixed charges replaced by charge tensors 
\cite{Pasquarello_97,Wilson_96} should be used for higher frequencies 
in order to describe properly the absorption peak positions and their relative 
intensities  across the IR vibrational band, 
but, in the FIR range, the rigid-ion model is adequate, as we have 
checked for our density functional theory-based tight-binding 
(DFTB) \cite{Porezag_95,Elstner_98} 
MD model of g-As$_2$S$_3$ \cite{Simdyankin_04}. 
%The temperature dependence (for $T\alt 100~\text{K}$) of $\alpha(\omega)$ for 
%$\omega/2\pi c \alt 1~\text{cm}^{-1}$ \cite{Strom_77} 
% can be possibly attributed  to excitations of two-level systems and/or 
% to highly anharmonic atomic modes and is not considered here. 

In ordered systems, 
the eigenmodes are phonons and static atomic charges do not fluctuate
and $C(\omega)$ is non-zero only 
for optic modes at the centre of the Brillouin zone. 
In disordered systems, structural disorder leads to a static charge 
transfer between atoms, i.e. to static disorder in the $q_i$, 
and to intrinsic disorder in the components 
of the vibrational  eigenmodes.
% which lose translational invariance. 
These two related sources of disorder are encoded in Eq.~(\ref{e2}) and  
are responsible for the peculiar behaviour of $C(\omega)$ 
in amorphous systems. 
In order to demonstrate this, we use two known features  about the structure 
of the vibrational eigenmodes in the FIR regime and about the 
distribution of atomic charges in glasses. 
First, the disordered eigenmodes in the low-frequency regime 
can be approximately expanded in plane waves 
(see  Ref.~\cite{Taraskin_00:PRB_IR1} for more detail) 
characterized by a wavevector ${\bf k}$ and unit 
polarization vector ${\bf {\hat p}}_{\bf k}$, i.e.   
${\bf e}_{i}(\omega) \simeq 
\sum_{\bf k}\sqrt{m_i/\overline{m}}~a_{\bf k}(\omega) {\bf {\hat p}}_{\bf k} 
e^{\text{i}{\bf k}(\omega)\cdot{\bf r}_i}$ with 
$\overline{m}=N^{-1}\sum_i m_i$ and ${\bf r}_i$ being the position vector for  
atom $i$ ($i=1,\ldots,N$). 
The distribution of coefficients $ a_{\bf k}(\omega)$ 
for low frequencies (at least below $\omega_{\text{IR}}$) has the  
shape of two relatively narrow peaks describing the disorder-induced 
hybridization between transverse and longitudinal acoustic phonons of 
the same frequency $\omega$, with  
the peak positions scaling linearly with frequency 
\cite{Taraskin_00:PRB_IR1,Taraskin_98:PhilMag}. 
Second, it has been found for an {\it ab-initio} MD model 
of g-SiO$_2$ that the atomic-charge fluctuations preserve approximately local 
charge neutrality 
within SiO$_4$ tetrahedra \cite{Pasquarello_97}. 
Similarly, we have found local charge neutrality to be obeyed within 
AsS$_3$ pyramids in a DFTB MD model of g-As$_2$S$_3$ \cite{Simdyankin_04}. 

These two observations lead to Eq.~(\ref{e1}).  
Assuming for simplicity that only a single plane wave with 
wave-vector ${\bf k}$ contributes to the disordered eigenmode 
(the presence of two peaks of finite widths in the  distribution 
$ a_{\bf k}(\omega)$ 
does not change qualitatively the results presented below, at least 
for $\omega \alt\omega_{\text{IR}}$),  
we can reduce the configurationally-averaged Eq.~(\ref{e2}) to the 
following form,  
$ \left\langle C(\omega)\overline{m}/C_0 \right\rangle \simeq 
N^{-1}  \langle \left|S\right|^2 \rangle$, 
where $S=\sum_i q_{i} e^{\text{i}{\bf k}\cdot{\bf r}_i}$. 
Further analysis is based on the following rather general 
properties of the random sum $S$: (i) if the values of $q_{i}$ 
are not random, then the orientationally averaged 
$S$
%, being proportional to the partial static 
% structure factors \cite{Elliott_90:book}, 
 scales linearly with $k$ for $k\to 0$ -  
the same holds for random but locally correlated values 
of $q_{i}$ (see below); and (ii) if the values of $q_{i}$ are random and 
uncorrelated, then $S \to \text{const.}$ for $k\to 0$. 
It is convenient to split the atomic charges  into two 
components, $q_i=q_{1i} +q_{2i}$, 
with $q_{1i}(\{ {\bf r}_i\})$ (depending on many atomic positions in a 
complicated fashion) representing uncorrelated charge components  so that 
$ \langle q_{1i}q_{1j} \rangle =
\langle q_{1i}\rangle\langle q_{1j} \rangle = 
\sigma_1^2 \delta_{ij}$ 
and e.g. 
$
\langle q_{1i} e^{\text{i}{\bf k}\cdot{\bf r}_j} \rangle 
\simeq 
\langle q_{1i}\rangle 
\langle e^{\text{i}{\bf k}\cdot{\bf r}_j} \rangle = 0
$.  
The random charges $q_{2i}$ satisfy local charge neutrality 
and can be imagined as resulting from charge transfers between 
nearest-neighbour atoms, i.e. 
$q_{2i} = \sum_{j\ne i}\Delta q_{ji} $, where $j$ runs over all 
 nearest neighbours of atom $i$ and $\Delta q_{ji}$ 
($=-\Delta q_{ij}$) is the 
charge transfer from originally neutral atoms $j$ to  $i$. 
In heteropolar crystals, the values of $\Delta q_{ji}$ 
are finite and not random.
In disordered systems, the values of $\Delta q_{ji}$ are distributed 
around mean values which do not necessarily coincide with those  
for their crystalline counterparts (see e.g. \cite{Pasquarello_97}). 
Such fluctuations in $\Delta q_{ji}$, and deviations of the means, are due to  
distortions in local structural units (e.g. 
the Si-O-Si bond angle in g-SiO$_2$) and the values of $\Delta q_{ji}$ are 
highly correlated on this length scale.  
The random charges $q_{1i}$ are not correlated with $q_{2j}$ and 
thus 
$\left\langle C(\omega)\overline{m}/C_0 \right\rangle = 
N^{-1}  
\langle|S_1|^2 +|S_2|^2 \rangle
$, where $S_n$ coincides with $S$ in which $q_{i}$ are replaced by $q_{ni}$.  
The first component, $N^{-1}\langle|S_1|^2\rangle = 
N^{-1}\sum_{ij} \langle q_{1i} q_{1j} \rangle 
\langle e^{\text{i}{\bf k}\cdot{\bf r}_{ij}}\rangle 
= \sigma_1^2$ 
is independent of $k$ and thus of $\omega$. 
The second component can be evaluated in the bond representation, in which 
$S_2 =
2\text{i}\sum_{(ij)}\Delta q_{ij}  e^{\text{i}{\bf k}\cdot\overline{\bf r}_{ij}} 
 \sin\left({\bf k}\cdot{\bf r}_{ij}/2\right)$ 
with ${\bf r}_{ij}={\bf r}_{j}-{\bf r}_{i}$ 
and  $\overline{\bf r}_{ij}=({\bf r}_{j}+{\bf r}_{i})/2$, 
and where the sum is taken over all the bonds $(ij)$ in the system. 
In the FIR regime, ${\bf k}\cdot{\bf r}_{ij} \ll 1$ and hence 
$S_2 \simeq k \sum_{(ij)}\Delta q_{ij}  e^{\text{i}{\bf k}\cdot
\overline{\bf r}_{ij}} 
(\text{i}{\hat{\bf k}}\cdot{\bf r}_{ij}) \equiv k\tilde{S}_2$. 
%with ${\bf k} = k{\hat{\bf k}}$. 
Consequently, the contribution to 
the coupling coefficient from correlated charges preserving 
local charge neutrality is  
$N^{-1}\langle|S_2|^2\rangle
\simeq k^2 \langle|\tilde{S}_2|^2\rangle
 \propto \omega^2 $ (assuming linear dispersion), 
where the function 
$ \langle |\tilde{S}_2|^2\rangle 
 = \text{const.}+\text{O}(k^2) 
$ depends on precise structural details of the material 
but does not depend on $k$ in the FIR range.  
%(which has been also checked 
%numerically, e.g. for a lattice model with positional disorder).  
Therefore, we have demonstrated that the frequency-independent 
part in Eq.~(\ref{e1}) is due to uncorrelated charge fluctuations while the 
quadratic frequency dependence results from correlated charge fluctuations 
satisfying local charge neutrality.  

The range of validity of Eq.~(\ref{e1}) is restricted at high frequencies 
 by the 
requirement that the disordered modes should have acoustic-like character 
with linear pseudo-dispersion. 
Generally speaking, for 
$\omega \agt \omega_{\text{IR}}$, 
the disordered modes contain many plane waves with a wavevector spread  
which is comparable with the mean value of $k$, and the dispersion law 
becomes modified, showing e.g. the onset of second sound for 
the longitudinal branch in g-SiO$_2$ \cite{Taraskin_00:PRB_IR1}. 
This means that the frequency dependence of $C(\omega)$ 
can be modified in a non-universal manner for $\omega \agt
 \omega_{\text{IR}}$,  as we actually found for g-SiO$_2$ and 
g-As$_2$S$_3$. 
Dynamical polarization effects can also become significant 
for $\omega \agt \omega_{\text{IR}}$ \cite{Gray-Weale_00}.  
Eq.~(\ref{e1}) is not restricted for low frequencies 
if possible low-frequency anharmonic effects are ignored \cite{Strom_77}.     

In order further to support the form  of the  
universal frequency dependence 
of the FIR coupling coefficient given by Eq.~(\ref{e1}), 
we give the following evidence: 
(i) experimentally derived data for $C(\omega)$ 
 in g-SiO$_2$ and g-As$_2$S$_3$, and  
numerical calculations of $C(\omega)$
for (ii)  classical MD models of 
g-SiO$_2$ with fixed charges and for $\alpha-$cristobalite 
(crystalline counterpart of g-SiO$_2$) with disordered charges and  
(iii) a DFTB MD model of g-As$_2$S$_3$ 
and of its crystalline counterpart, orpiment, with disordered Mulliken 
charges. 

\begin{figure}[ht] % 
%\vskip2truecm 
%\centerline{\includegraphics[width=8cm]{./Figs/fig1}}  
\centerline{\includegraphics[width=8cm]{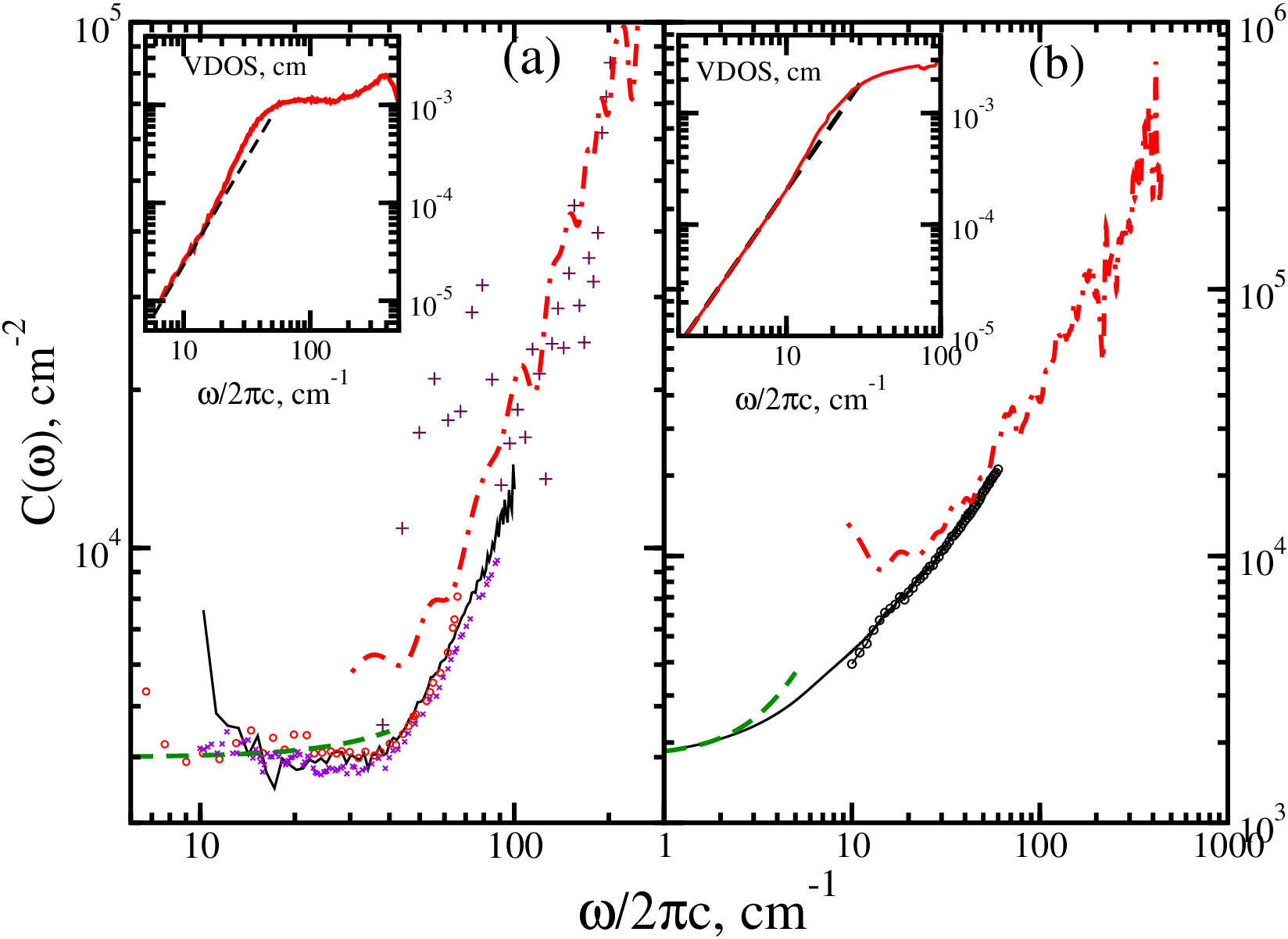}}  
\caption{(Color online) 
Experimental TDS frequency dependence of the absorption 
coupling coefficient in the FIR range for 
(a) g-SiO$_2$ and (b) g-As$_2$S$_3$ (solid lines). 
Other experimental data from Ref.~\cite{Grischkowsky_90} ($\circ$), 
Ref.~\cite{Ohsaka_98} ($\times$), Ref.~\cite{Henning_97} ($+$) in 
(a) and Ref.~\cite{Ohsaka_94}($\circ$) in (b) are shown for comparison.   
The dot-dashed lines represent the numerical data obtained from MD models. 
The dashed lines show the fits of the experimental data by Eq.~\ref{e1} 
with $A=4000~\text{cm}^{-2}$, $B=0.3~\text{cm}^{-1}$ for g-SiO$_2$ and 
$A=1780~\text{cm}^{-2}$, $B=75~\text{cm}^{-1}$ for g-As$_2$S$_3$. 
The insets show the VDOS used to extract  $C(\omega)$. 
The dashed line in the insets show the 
Debye VDOS, with the Debye temperature being $330$ K for  
g-SiO$_2$ and $164$ K for g-As$_2$S$_3$. 
  }  
\label{f1} 
\end{figure}

Experimental data for the FIR absorption coefficient 
of  spectrosil WF  
g-SiO$_2$ and ultra-pure g-As$_2$S$_3$ at room temperature 
have been obtained by THz-TDS  
\cite{Grischkowsky_90} in the 
frequency range $1- 130~\text{cm}^{-1}$ 
 using a TeraView  TPI Spectra 1000 transmission spectrometer 
(experimental details will be given elsewhere). 
Experimental data for $\alpha(\omega)$ have been divided 
by the experimental VDOS obtained from inelastic neutron scattering 
\cite{Fabiani_05,Isakov_93} (see the insets in Fig.~\ref{f1}) 
and $C(\omega)$ thus obtained is shown 
in Fig.~\ref{f1}. 
Our data agree well with other experimental FIR  
data obtained by THz-TDS or other techniques 
\cite{Grischkowsky_90,Hutt_89,Ohsaka_94,Henning_97,Ohsaka_98}
 and $C(\omega)$ can be well fitted by Eq.~(\ref{e1}) in the frequency 
range of its applicability  
(see the dashed lines in Fig.~\ref{f1}).

%------------------------------------------------------------------------ 
%           Figure 1 
%------------------------------------------------------------------------ 
%-----------------------------------------------------------------------

\begin{figure}[ht] % 
%\vskip20pt
%\centerline{\includegraphics[width=8cm]{./Figs/fig2}}  
\centerline{\includegraphics[width=8cm]{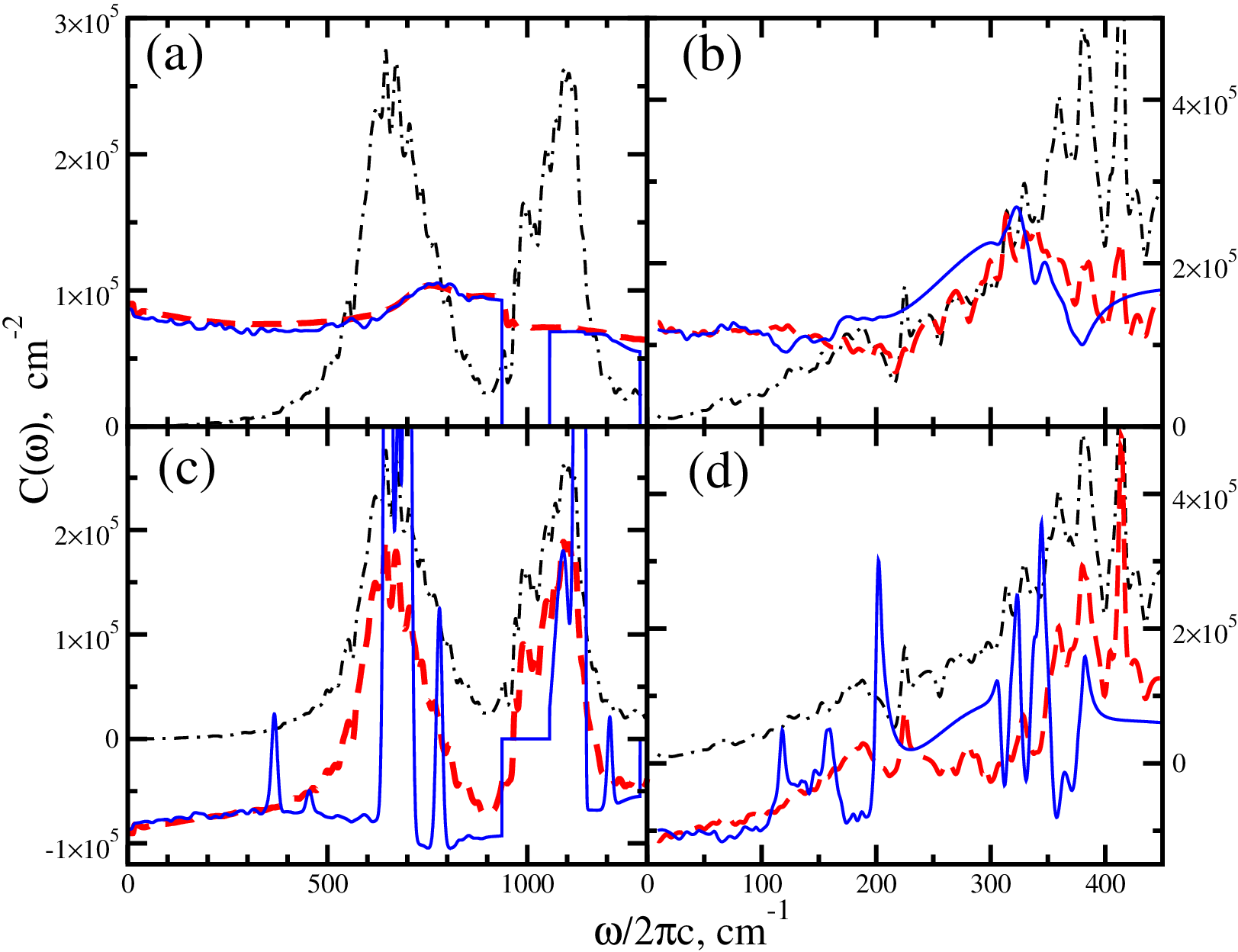}}  
\caption{(Color online) 
Incoherent (a),(b) and coherent (c),(d) contributions to 
the coupling coefficients  for MD models  
of g-SiO$_2$ and $\alpha-$cristobalite  (a), (c)
and of g-As$_2$S$_3$ and orpiment (b), (d)   
to the total coupling coefficient (dot-dashed line). 
Solid lines represent crystalline models and dashed lines 
are for glassy ones. 
} 
\label{f2} 
\end{figure}

We have also calculated $C(\omega)$ for a classical MD model 
of g-SiO$_2$\cite{Taraskin_99:PRB}  
and a DFTB MD model of g-As$_2$S$_3$ \cite{Simdyankin_04} (see the dot-dashed 
lines in Fig.~\ref{f1}).  
The classical MD model of g-SiO$_2$ has fixed atomic charges 
$q_{\text{Si}}=2.4$ and $q_{\text{O}}=-1.2$,  and 
thus $C(\omega)$ should not contain the frequency-independent 
component, $\langle|S_1|^2\rangle$, and should be proportional to 
 $\omega^2$ for $\omega/2\pi c \equiv \tilde{\omega} 
\alt \tilde{\omega}_{\text{IR}}  
\simeq 35~\text{cm}^{-1}$ solely due to positional disorder.  
This region, unfortunately, is not available in our MD model due 
to prominent finite-size effects below $50~\text{cm}^{-1}$. 
However, in the range $50 - 300~\text{cm}^{-1}$, 
we have found $C(\omega) \propto \omega^{\beta}$ with $\beta \simeq 2$ 
which can be due to the existence of a peak-shaped spectral 
density and a linear pseudo-dispersion law in this frequency range 
\cite{Taraskin_00:PRB_IR1}.  
The DFTB MD model of g-As$_2$S$_3$  incorporates both charge 
and positional types of disorder and thus can show the crossover from 
a quadratic frequency dependence to the plateau behaviour with decreasing 
frequency (but this is also unavoidably masked by 
finite-size effects for $\tilde{\omega} \alt 20~\text{cm}^{-1}$) 
as seen in Fig.~\ref{f2}(b). 
Between the BP frequency, 
$\tilde{\omega}_{\text{BP}} \simeq 15~\text{cm}^{-1}$ \cite{Sokolov_93} 
and $200~\text{cm}^{-1}$, 
we have found $C(\omega)\propto \omega^\beta$, with $\beta\simeq 1.7$,  
which differs from that found for g-SiO$_2$ and 
might be due to non-linear pseudo-dispersion in this frequency range in 
g-As$_2$S$_3$ and possible dynamic-polarization effects 
\cite{Gray-Weale_00}.

%------------------------------------------------------------------------ 
%                       Figure 2
%------------------------------------------------------------------------ 

%------------------------------------------------------------------------------------------------------------------------------

\begin{figure}[ht] % 
%\vskip20pt
%\centerline{\includegraphics[width=8cm]{./Figs/fig3}} 
\centerline{\includegraphics[width=8cm]{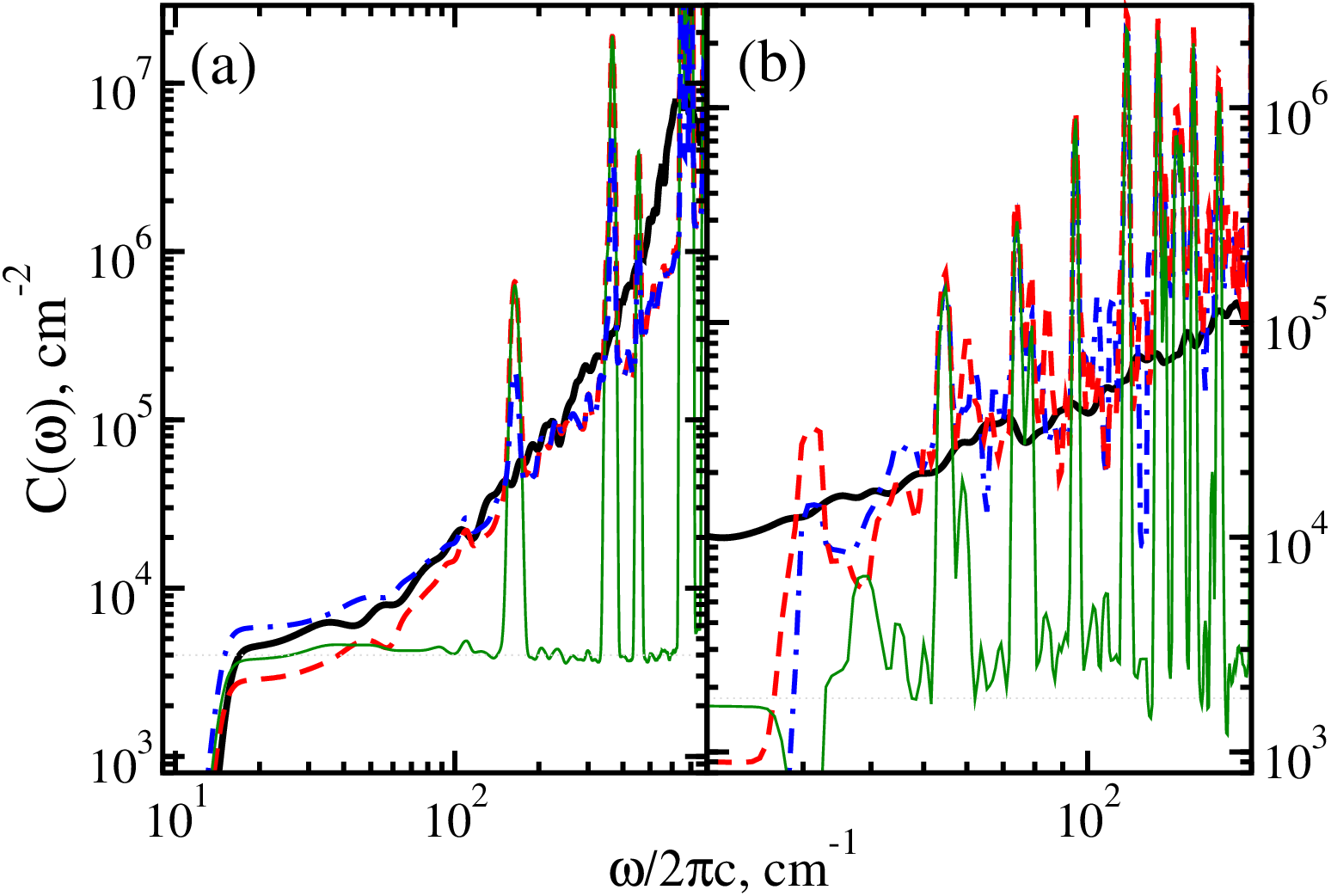}} 
\caption{(Color online) 
Frequency dependence of $C(\omega)$ for 
MD models of $\alpha-$cristobalite (a) and orpiment (b) with
uncorrelated (thin solid lines), correlated (dot-dashed lines)
and both types of charge disorder (dashed lines).
The uncorrelated charges were drawn from normal distributions
with standard deviations,
$\sigma_{\text{Si}}/q_{\text{Si}} =
\sigma_{\text{O}}/|q_{\text{O}}|=0.04$
and $\sigma_{\text{As}}/q_{\text{As}}=
\sigma_{\text{S}}/|q_{\text{S}}|=0.06$, and zero  means. 
The correlated charges were taken from a normal distribution with
 mean values and standard deviations of:
$\langle q_{2i,\text{Si}}\rangle=2.4$,
$\langle q_{2i,\text{O}}\rangle=-1.2$,
$\langle q_{2i,\text{As}}\rangle=0.56$,
$\langle q_{2i,\text{S}}\rangle=-0.37$ and
$\sigma_{2,\text{Si}}/\langle q_{2i,\text{Si}}\rangle=0.6$,
$\sigma_{2,\text{As}}/\langle q_{2i,\text{As}}\rangle=0.6$.
The random values of $q_{2i}$ have been compensated by
$-q_{2i}/Z$ placed on $Z$ nearest neighbours in order to maintain 
local charge neutrality.
The parameters of the distributions for uncorrelated charges were chosen 
to fit the experimental data for the plateau value (parameter $A$ in 
Eq.~(\ref{e1})),  and for correlated
charges  to fit
the frequency dependence of $C(\omega)$ in the glassy MD
models (solid lines).
}
\label{f3} 
\end{figure}

In order to unmask finite-size effects and reveal the role 
of charge fluctuations in the 
frequency dependence of $C(\omega)$, 
we argue that all the universal features of 
$C(\omega)$ can be seen in lattice models 
of corresponding crystalline counterparts incorporating 
correlated and uncorrelated charge disorder. 
Many properties of disordered systems are 
rather similar to those of their crystalline counterparts 
(see e.g. \cite{Taraskin_00:PRB_IR1,Simdyankin_00}). 
$C(\omega)$ shares such a similarity as well. 
In order to see this, it is convenient to split the    
 expression for $C(\omega)$ into 
incoherent (self-atom) and coherent (correlated-atom) components, 
$C(\omega)= C^{\text{incoh}}(\omega)+C^{\text{coh}}(\omega)$, 
where e.g. 
$C^{\text{incoh}}(\omega)= C_0 \sum_i (q_i^2/m_i){\bf e}^2_{i}(\omega)$. 
% and  
%$C^{\text{coh}}(\omega)= 
%C_0 \sum_{i\ne j} (q_iq_j/\sqrt{m_im_j})
%{\bf e}_{i}(\omega)\cdot {\bf e}_{j}(\omega)$.  
These quantities, calculated both for MD models of 
amorphous solids and their crystalline 
counterparts, show striking similarities    
(cf. the solid and dashed lines in Fig.~\ref{f2}(a,b)), 
which  suggests that 
the frequency dependence of $C(\omega)$ 
in amorphous systems can be mimicked  
by introducing  correlated and uncorrelated charge disorder 
 into crystalline systems (see also \cite{Schlomann_64}).  
Indeed, we have found that 
uncorrelated disorder on its own, as expected, results 
in a frequency-independent coupling constant 
(see the thin solid lines in Fig.~\ref{f3}),  
the value 
of which depends on the variance of the distribution of $q_{1i}$, 
i.e. $\sigma_1^2$.   
Correlated disorder on its own gives rise to the same 
frequency dependence (see the dashed lines in Fig.~\ref{f3}) 
of $C(\omega)$ found 
above the BP in the MD models (cf. the dashed and solid lines 
in Fig.~\ref{f3}). 
The slope of the curves does not depend on the variance, $\sigma_2^2$, 
which only influences the intercept on the vertical axis. 
Incorporation of both types of disorder 
(see the dot-dashed lines in Fig.~\ref{f3}), with the 
plateau value coinciding with the parameter $A$ found from the fit 
of Eq.~(\ref{e1}) to experimental data,   
reveals rather well the frequency dependence of $C(\omega)$ 
found in the MD models above the BP    
(cf. the dot-dashed and solid lines in Fig.~\ref{f3}). 
Such a fit allows the widths of 
the distribution of uncorrelated charges, $\sigma_1 =\sqrt{A\overline{m}/C_0}$, 
to be estimated, 
$\sigma_{1,\text{SiO}_2} \simeq 0.06~e$ and 
$\sigma_{1,\text{As}_2\text{S}_3} \simeq 0.12~e$; these values 
are in accord with those expected from the relative ionicities 
of the two materials. 
Below $50~\text{cm}^{-1}$ (for g-SiO$_2$) and  
$20~\text{cm}^{-1}$ (for g-As$_2$S$_3$), the simulation data 
are not reliable due to finite-size effects. 
However, extrapolation to lower frequencies (larger sizes) is 
rather straightforward for lattice models with charge disorder. 
For example, in the case of g-SiO$_2$, the contribution from  
correlated charge fluctuations (the dashed line in Fig.~\ref{f3}(a)) 
becomes insignificant below the BP 
in comparison with the dominant  frequency-independent contribution from 
uncorrelated charge fluctuations (the thin solid line in Fig.~\ref{f3}(a) 
which should be extended to lower frequencies for larger models). 
Therefore, the sum of correlated and uncorrelated charge fluctuations should 
result in a crossover to a plateau  for frequencies below the BP, 
i.e. as in the experimentally observed behaviour of $C(\omega)$ 
(see Fig.~\ref{f1}).  

%------------------------------------------------------------------------ 
%                       Figure 3
%------------------------------------------------------------------------ 

%------------------------------------------------------------------------------------------------------------------------------

In conclusion, we have presented an explanation for the universal 
frequency dependence of the coupling coefficient 
for far-infrared photons with atomic vibrations. 
The coupling coefficient below the Ioffe-Regel crossover has two components. 
One is frequency independent and is due to uncorrelated static charge 
fluctuations caused by medium- and 
long-range structural irregularities. 
This results in the quadratic frequency dependence of the absorption
coefficient frequently observed in disordered crystals and glasses.  
The other contribution  depends quadratically on frequency and is caused 
by structural disorder on the short-range (interatomic) scale,  
leading to static correlated charge fluctuations obeying local charge 
neutrality within structural units. 
For two glasses studied, g-SiO$_2$ and g-As$_2$S$_3$, we can conclude 
that, in  g-SiO$_2$, uncorrelated charge fluctuations 
dominate through the whole frequency range below the Ioffe-Regel 
crossover  (or boson peak)  and result in a 
frequency-independent absorption coupling 
coefficient there. 
In contrast, in g-As$_2$S$_3$, uncorrelated charge fluctuations are less 
pronounced and correlated charge fluctuations preserving local charge 
neutrality become appreciable even below the BP and thus 
the absorption coupling coefficient exhibits an onset 
to an $\omega^2$-dependence in this frequency range. 
 
SIS thanks the Newton Trust, Cambridge and 
JRN is grateful to the International Materials Institute for 
New Functionality in Glass for financial support.  
We are grateful to Z. Hajnal for the PhonIR package used in 
vibrational analysis of the DFTB data, 
to Prof. U. Buchenau for discussions and supplying us with 
inelastic neutron-scattering data for v-SiO$_2$ and to 
Prof. M.F. Churbanov for the chalcogenide samples.  

%\bibliography{../Archive/archive_snt}

\end{document}